\documentclass[11pt]{iopart}

\usepackage{iopams}

\usepackage{cases}
\usepackage{makeidx}
\makeindex
\usepackage {graphicx}
\usepackage{color}

\def\>{\right\rangle}
\def\<{\left\langle}
\def\be{\begin{equation}}
\def\ee{\end{equation}}
\def\ba{\begin{array}{lll}}
\def\ea{\end{array}}

\def\beq{\begin{eqnarray}}
\def\eeq{\end{eqnarray}}
\begin{document}

\title{Multiple quasiparticle Hall spectroscopy investigated with a resonant detector}
\author{\bf{D Ferraro}$^{1,2,3}$, M Carrega$^{4}$, A Braggio$^{5}$, and M Sassetti$^{6,5}$}
\address{$^1$ Universit\'e de Lyon, F\'ed\'eration de Physique Andr\'e Marie Amp\` ere, CNRS - Laboratoire de Physique de l'Ecole Normale Sup\'erieure de Lyon, 46 All\'ee d'Italie, 69364 Lyon Cedex 07, France}
\address{$^2$ Aix Marseille Universit\'e, CNRS, CPT, UMR 7332, 13288 Marseille, France}
\address{$^3$ Universit\'e de Toulon, CNRS, CPT, UMR 7332, 83957 La Garde, France}
\address{$^4$ NEST, Istituto Nanoscienze-CNR, and Scuola Normale Superiore, I-56126, Pisa, Italy}
\address{$^5$ SPIN-CNR, Via Dodecaneso 33, 16146 Genova, Italy}
\address{$^6$ Dipartimento di Fisica, Universit\`a di Genova, Via Dodecaneso 33, 16146 Genova, Italy}

\begin{abstract}
We investigate the finite frequency (f.f.) noise properties of edge 
states in the quantum Hall regime.
We consider the measurement scheme of a resonant detector coupled to a 
quantum point contact in the weak-backscattering limit.
A detailed analysis of the difference between the ``measured'' noise, 
due to the presence of the resonant detector, and the symmetrized f.f. 
noise is presented.
We discuss both the Laughlin and Jain sequences, studying the tunnelling excitations in these hierarchical models.
We argue that the measured noise can better distinguish between the different excitations in the 
tunnelling process with respect to the symmetrized f.f. counterpart in an experimentally relevant range of parameters.
Finally, we illustrate the effect of the detector temperature on the 
sensibility of this measure.\end{abstract}

\section{Introduction}

One of the most intriguing properties of fractional quantum Hall liquids is the existence of fractional charges intrinsically connected to the strong interacting nature of the system.~\cite{DasSarma97} Shot noise~\cite{Schottky18} measurements in a quantum point contact (QPC) geometry has played a fundamental role in detecting these charges. Indeed, the zero frequency current-current correlation, in the weak-backscattering regime, is predicted to be proportional to the induced backscattering current via the fractional charge associated to the tunnelling excitation between the opposite edges of the Hall bar.~\cite{Wen95,Kane94,Kane95} Clear experimental signature of this fact has been reported for the Laughlin sequence~\cite{Laughlin83} with filling factor $\nu=1/(2n+1)$, with $n \in \mathbb{N}$, and with measured fractional charges $e^{*}= e/ (2n+1)$ ($e$ the electron charge) in agreement with theoretical predictions.~\cite{Kane94b,DePicciotto97, Saminadayar97,Reznikov99}  

More involved is the situation concerning the Jain sequence~\cite{Jain89}, with $\nu=p/(2np+1)$, where edges in general have a composite nature with fundamental charge $e^{*}= (\nu/p) e$. Experiments show cross-over of the measured effective charge evolving between different values while varying the temperature.~\cite{Chung03, Bid09, Dolev10} This peculiar behaviour has been explained in terms of competition of different tunnelling excitations, the so-called $m$-agglomerates, with charge $m e^{*}$, whose dominance depend on the considered energy scale.~\cite{Ferraro08, Ferraro10, Ferraro10b, Carrega11, Wang13, Shtanko13, Smits14}
Nevertheless, zero frequency noise may be not enough in order to extract in a univocal way the values of the fractional charges when many of them contribute, with comparable weight to QPC transport. This is a consequence of the fact that zero frequency noise probes a weighted average between different charges when these are present.

A possible way to overcome this limitation is to look at the finite frequency (f.f.) properties.~\cite{Blanter00} In particular, for quantum Hall QPC transport, the f.f. noise is predicted to show resonances in correspondence of Josephson frequencies, which are proportional to the fractional charges.~\cite{Rogovin74,Chamon95, Chamon96, Dolcini05, Bena06, Bena07, Carrega12, Ferraro12}

From a theoretical point of view the first investigations focussed on the f.f. symmetrized noise that coincides with the classical definition of noise spectrum.~\cite{Chamon95, Chamon96} Here the presence of an $m$-agglomerate tunnelling may be revealed, at extremely low temperatures, with the presence of peaks or dips in the noise spectrum occurring at frequency $\omega\backsim m e^{*} V/\hbar$. However, for frequencies in the range $\hbar\omega\gg k_B T$, one needs to take care about quantum effects (quantum noise). At that short timescales, indeed, the classical definition of f.f. symmetrized noise is not proper anymore to explain the general measurement scheme.~\cite{Lesovik97,Gavish00,Gavish03,Bednorz13} One needs to carefully consider the circuit of detection identifying which quantity is effectively probed. In this context Lesovik and Loosen in Ref.~\cite{Lesovik97} introduced a model of a resonant circuit as the prototypical scheme for f.f. noise measurement. Soon after this seminal paper, other detection schemes have been proposed.~\cite{Gavish00,Aguado00,Creux06,Zazunov07,Gabelli08, Chevallier10}  In particular it  has been shown that the measured quantity for the LC detector setup can be expressed in terms of the non-symmetrized f.f. noise which reflects the emission and adsorption contributions of the active system (in our case the QPC) under investigation. The non-symmetrized noise has been considered in literature for different systems as the ultimate resource of information of quantum noise properties.
This quantity have been also considered in non-interacting systems~\cite{Aguado00} and for interacting electron in one dimension~\cite{Sukhorukov01, Johansson02, Schoelkopf03, Deblock03, Engel04, Hekking06, Billangeon06, Levchenko08, Cottet08, Safi08, Safi09}. 
An analysis of the f.f. noise measuring process, discussing the issue of measurability of the zero point fluctuation and the role of the amplifiers in quantum detection, has been diffusely done in the literature.~\cite{Gavish00,Gavish03,Bednorz13} 

In this paper we will consider the detector model of Ref.~\cite{Lesovik97}, coupled with a QPC in the fractional Hall regime, investigating the f.f. detector output power.  This realistically measurable noise will be analyzed at fixed frequency $\omega$ as a function of bias $\omega_0=e^*V/\hbar$. We will assume that the temperature  $T_c$ of the detector could be controlled and kept different from the QPC one ($T$). We will mainly consider the quantum limit for the detector, $\hbar\omega\gg k_B T_c$, where the output power is proportional to the non-symmetrized noise. The QPC, will be investigated by scanning the bias in and out of equilibrium (shot noise limit $e^*V\gg k_B T$). These limits represent the best conditions to extract information about fractional quasiparticles (qps), in particular their charge $me^*$ and their scaling properties. Indeed, in this parameter range the measured noise, returns precisely the $m$-agglomerate tunnelling rate. 

Firstly,  we will analyze the well known case of non-interacting Fermi liquid ($\nu=1$) and Laughlin ($\nu=1/3$)  to show some important and useful properties of the measurement setup. Differently from what usually considered in other theoretical papers, where the noise is shown at finite bias as a function of the frequency, here we will discuss the opposite case in which the bias is moved at fixed frequency. This allows us to be closer to realistic experimental situations representing by far the simplest measurement protocol for the system. We will discuss in details the advantages of considering this measurement scheme in comparison to the simpler symmetrized noise. Moreover, the detector response will give the unique possibility to selectively address the emission contribution of QPC noise or its adsorptive part only by acting on the detector temperature. We also discuss the range of the detector temperatures in order to access the non-symmetrized noise contributions. It is convenient that $T_{c}$ is smaller than the considered frequency.
In the final part of the paper we apply the previous concepts to the measurement of multiple qps for two values of the Jain sequence ($\nu=2/5$ and $\nu= 2/3$). 

In all cases we demonstrate how this setup is able to clearly address the different qps contributions separately and to quantitatively validate the hierarchical edge state models.~\cite{WenZee92}

The paper is organized as follows. In section \ref{theomodel} we resume the edge state models for the Laughlin (single mode edge state $\nu=1,1/3$) and Jain sequence (composite edge state $\nu=2/5,2/3$). We clarify the scheme of detection of the quantum Hall QPC, in the weak-backscattering regime and weakly coupled to a resonant LC circuit. We discuss the symmetrized $S_{sym}$ and non-symmetrized noise $S_\pm$ for the QPC backscattering current and its connection with the measured output signal from the LC detector, the measured noise $S_{meas}$. We show the relation between the latter quantity with and previous f.f. noise correlators and we comment on some useful limits. In particular we show that $S_{meas}$ contains the non-symmetrized noise as a limit case but it can return more general information by addressing other regimes. Finally
we show how these quantities can be connected to the qps tunnelling rates and the physical properties of the tunnelling excitations. In section \ref{results} we compare the f.f. symmetrized noise and the measured noise for the Laughling case, introducing the essential concepts which will be useful in the investigation of the Jain sequence. Finally for the Jain sequence we show how this detection scheme is helpful in the investigation of the multiple quasiparticle tunnelling for composite edge models.
 
\section{Theoretical model}
\label{theomodel}
In this section we briefly remind the effective field theories describing edge states in the fractional quantum Hall effect (FQHE) for filling factors in the Laughlin~\cite{Laughlin83} and the Jain~\cite{Jain89} sequences.
We recall only the main results referring to the literature for a more detailed discussion~\cite {Wen95,Kane94, Kane95, Ferraro08, Ferraro10, Ferraro10b}.

\subsection{Laughlin and Jain sequences}
In the field theoretical description~\cite{WenZee92}, edge states of the Laughlin~\cite{Laughlin83} and 
Jain~\cite{Jain89} sequences are expressed in terms of 1D chiral bosonic modes. For the filling 
factors $\nu=p/(2np+1)$ the edge states satisfy an hidden $SU(|p|)$ symmetry involving $|p|$ different modes.~\cite{WenZee92} One can express these states in terms of a single charged mode and $|p|-1$ neutral ones.~\cite{Kane95} Hereafter, for simplicity, we will consider the cases with $|p|\leq 2$ but many of the results can be easily generalized to other cases.  

For the Laughlin sequence ($|p|=1$) one has a single edge charged chiral mode. 
The case $n=0$ corresponds to the integer quantum Hall state $\nu=1$, a chiral Fermi liquid that represents the 
non-interacting limit. Instead, $n=1$ describes the most stable fractional quantum Hall phase the $\nu=1/3$.  The Lagrangian density associated to the charged mode $\varphi_{c}$ reads
\be
\mathcal{L}_c=- \frac{1}{4 \pi \nu_c}  \partial_{x}\varphi_{c}  ( \partial_{t} +v_{c} \partial_{x} ) \varphi_{c},
\label{lag_wen_c}
\ee
where $\nu_c=\nu$ and $v_c $ is the mode velocity. The charged mode is related to the electron particle 
density along the edge through $\rho=\partial_x\varphi_c/2\pi$. 

For $n=1$ one has also composite edge, for example $\nu=2/5$ with $p=2$ and $\nu=2/3$ with $p=-2$ that contain an additional neutral mode.~\cite{Kane95}  In particular, it has been demonstrated that the low energy effective theory for $\nu=2/5$ may be conveniently described in terms of a co-propagating neutral mode and, instead, a counter-propagating one is required for $\nu=2/3$.~\cite{Kane94, Kane95} Together with the charged part, see Eq.(\ref{lag_wen_c}), the Lagrangian density of these states presents also a neutral sector contribution $\varphi_{n}$ 
\be
\mathcal{L}_n=-\frac{1}{4 \pi \nu_{n}}  \partial_{x}\varphi_{n}  (\mathrm{sgn}(p) \partial_{t} +v_{n} \partial_{x} ) \varphi_{n},
\label{lag_wen_5}
\ee
where $\nu_{n}=2$ and $\mathrm{sgn}(p)$ encodes  the co- or counter-propagating property of the neutral mode. 
In general one could expect that the neutral mode velocity $v_n$ satisfies $v_{n}\ll v_{c}$ since the charge 
mode is more strongly affected by Coulomb interactions.~\cite{Levkivskyi08}

The bosonic fields satisfy the general commutation relations (commutators not explicitly written are zero) 
\beq
\label{commutazione_wen_5}
\left[\varphi_{c}(x) , \varphi_{c}(x') \right]& =& i \pi  \nu_{c} \mathrm{sign}(x-x')\\
\left[\varphi_{n}(x) , \varphi_{n}(x') \right]& =& i \pi \mathrm{sgn}(p) \nu_{n} \mathrm{sign}(x-x'),
\eeq
which determine the statistical properties of the excitations. 
The annihilation operator of these excitations, obtained by the monodromy~\cite{Froelich97, Ino98} requirement, is defined by
\be
\label{op_wen_5}
\Psi_\nu^{(m,j)} (x) = \frac{\mathcal{F}^{(m,j)}}{\sqrt{2 \pi a}} e^{i \left( \frac{m}{|p|} \varphi_c (x) + \frac{j}{2} \varphi_n (x) \right)},
\label{vertex_jain}
\ee
with $m$ and $j$ integers with the same even/odd parity and $a$ a short-length cut-off. In the above equation $\mathcal{F}^{(m,j)}$ indicates the Klein factor~\cite{Guyon02, Martin05}. Note that for the Laughlin sequence $\nu=1/(2n+1)$ the neutral component is absent and $j\equiv 0$. The index $\nu$ in
the operator reminds the physical properties of the excitations (charge and statistics) which are tightly connected to the filling factor.  

For example, one can show that the charge associated to the excitation, described by $\Psi_\nu^{(m,j)}(x)$, is an integer multiple $me^*$ of the fundamental charge $e^*=\nu e/|p|$, justifying the name 
of $m$-agglomerate that we will use throughout all the paper. The fundamental charge, for the cases we will consider, are, the electron charge $e^{*}=e$ for the chiral Fermi liquid ($\nu=1$), $e^{*}=e/3$ for filling factors $\nu=1/3$ and $\nu=2/3$ and $e^{*}=e/5$ 
for $\nu=2/5$. 

An important property of the operators $\Psi_\nu^{(m,j)}$ is their scaling dimension $\Delta_\nu^{(m,j)}$ defined as the imaginary time scaling of the two point Green function $\langle \Psi_\nu^{(m,j)} (x,\tau)\Psi_\nu^{(m,j)\dagger} (x,0)\rangle\propto|\tau|^{-2\Delta_\nu^{(m,j)}}$ for $|\tau|\to \infty$ (low energy limit). For the excitation in Eq. (\ref{vertex_jain}) it reads
\be
\Delta^{(m,j)}_\nu=\Delta^{(m,j)}_{\nu,c}+\Delta^{(m,j)}_{\nu,n}= \frac{1}{2}\left(g_{c} \nu_{c}\frac{m^{2}}{p^2}+ g_{n} \nu_{n} \frac{j^{2}}{4}\right),
\label{scaling_jain}
\ee
where we have introduced $\Delta^{(m,j)}_{\nu,c}$ ($\Delta^{(m,j)}_{\nu,n}$) to identify the charged (neutral) contribution. The scaling dimension determines the energy dependence of the tunnelling density of state for any excitation $\Psi_\nu^{(m,j)}$. 
 In the previous equation, for sake of generality, we have introduced 
renormalization coefficients $g_{c}$ and $g_{n}$ which take into account possible mechanisms of modes interaction with external degrees of freedoms.~\cite{Braggio12} These mechanisms have been recently considered in order to better match the theory with experimental observations.~\cite{Ferraro08, Ferraro10, Ferraro10b,Carrega11}. 
 
According to the previous formula, one can identify at low energies the most relevant excitations corresponding to the operators $\Psi_\nu^{(m,j)}$ that have the lowest scaling dimension. 

For the Laughlin case it is the single-quasiparticle (qp) $\Psi_\nu^{(1,0)}$ that dominates the tunnelling in the low energy limit. Namely the electron $e$ (for $\nu=1$) and the fractionally charged excitation with $e^*=\nu e$ for $\nu=1/(2n+1)$ ($n>0$).

For the composite edge states ($|p|= 2$) the situation is more intricate.\footnote{For the state in the Jain sequence with $|p|=2$: the single-qp has $\Delta_{\nu,n}^{(1)}=g_n/4$ and $\Delta_{2/5,c}^{(1)}=g_c/20$ and  $\Delta_{2/3,c}^{(1)}=g_c/12$. The $2$-agglomerate instead does not contain any neutral component (even $m$), namely $\Delta_{\nu,n}^{(2)}=0$, and the scaling dimension of the charge component are $\Delta_{2/5,c}^{(2)}=g_c/5$ and $\Delta_{2/3,c}^{(2)}=g_c/3$.} 
The two most relevant excitations are the single-qp $e^*=\nu e/|p|$ ($m=1$) and the $2$-agglomerate $2e^*$ ($m=2$).
Their dominance on the transport depends on temperature, voltage, relative opacity of the tunnelling barrier and renormalization parameters $g_c$ and $g_n$.~\cite{Chung03,Bid09,Ferraro08, Ferraro10, Ferraro10b,Carrega11} 

For example, for $\nu=2/5$ (with $g_c=g_n=1$, i.e. unrenormalized case) the $2$-agglomerate $\Psi_{2/5}^{(2,0)}$ is dominant over the single-qp
$\Psi_{2/5}^{(1,1)}$, at low energy, and only at intermediate energies the single-qp becomes dominant.~\cite{Chung03,Bid09,Ferraro08} This fact is even more relevant for $\nu=2/3$, where in the unrenormalized case the single-qp $\Psi_{2/3}^{(1,1)}$ and the $2$-agglomerate $\Psi_{2/3}^{(2,0)}$ have the same scaling dimension, i.e. $\Delta_{2/3}^{(1,1)}=\Delta_{2/3}^{(2,0)}$. Moreover, as shown in experiments~\cite{Chung03, Bid09, Dolev10}, some renormalization mechanism for the exponent seems to be necessary in order to reconcile the hierarchical theory predictions with experimental observations.~\cite{Ferraro08, Ferraro10, Carrega11}. In the following, for sake of simplicity, we will consider only the bare theory with $g_c=g_n=1$. In order to further simplify the notation, we will suppress the neutral mode index for 
the composite edge model indicating the generic $m$-agglomerate as $\Psi_\nu^{(m)}$. Keeping only the 
most dominant excitations in the tunelling processes.
The use of the general concept of $m$-agglomerate opens the possibility to easily extend our results to all the other hierarchical edge models  and, eventually to some non-Abelian models such as $\nu=5/2$.~\cite{Carrega11,Carrega12}

\begin{figure}[ht]
\centering
\includegraphics[scale=.55]{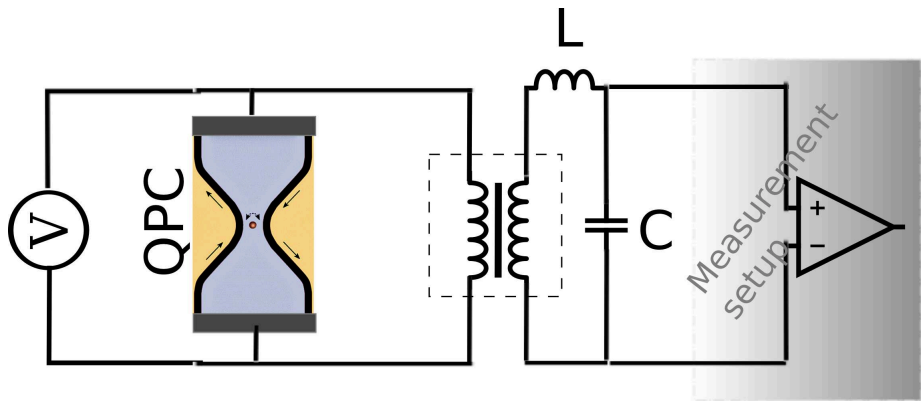}
\caption{Schematic view of an Hall bar (blue) with a QPC coupled with LC detector. A bias voltage $V$ is applied to the QPC and $m$-agglomerate excitations can tunnel between the edges. The two circuits are impedance matched via a coupling circuit (inside dashed line).}
\label{fig_A}
\end{figure}

\subsection{Backscattering current fluctuations in a QPC geometry}
Our task is to study the f.f. backscattering current-current fluctuations in a single 
QPC geometry coupled to the detector set-up shown in Fig.~\ref{fig_A}.
A similar prototype setup was proposed by Lesovik and Loosen in Ref.~\cite{Lesovik97} for a non-interacting junction: the QPC is subjected to a bias voltage V and
coupled to a resonant LC circuit, the detector (with frequency $\omega=\sqrt{1/LC}$), via an impedance matching circuit 
 (inside the dashed line in the figure). We are not interested in the precise form of the coupling $K$ between QPC and detector and demand to literature for further details~\cite{Lesovik97, Chevallier10}, assuming an equivalent electrical coupling $K\ll1$ between QPC and detector.\footnote{The LC detector is electrically connected to the measurement  setup (grey shaded area). This small signal extraction may be modelled as a small resistance that we assume to not degrade the ideal LC circuit quality factor.}
 
The frequency $\omega$ is kept fixed assuming a very high quality factor of the detector. Since edge states have a chiral nature one could also consider a more general scheme 
with multi terminal setup, as is usually done in quantum 
Hall bars. This has been discussed in literature~\cite{Zamoum13} and the following results can be  easily generalized to those cases. However, to make the discussion as simple as possible, we focus on the two terminal geometry only. 

We consider a quantum Hall point contact where the tunnelling processes between 
the edges are weak (weak-backscattering limit). Thus we will treat separately the 
contribution of the different $m$-agglomerate excitations to the transport properties. The point-like tunnelling of a generic $m$-agglomerate between the upper ($+$) and the lower ($-$)  edge can be described through the tunnelling Hamiltonian
\be
\label{tun}
\hat{{\cal H}}^{(m)}_{T}= t_{m}\  \Psi_{\nu,+}^{(m)}(0){\Psi^{(m)}_{\nu,-}}^{\dagger}(0)+h.c.\ ~,
\ee
where $t_{m}$ is the tunnelling amplitude associated to the $m$-agglomerate excitation of up (down) edge with annihilation operator $\Psi^{(m)}_{\nu,+}$ ($\Psi^{(m)}_{\nu,-}$).\footnote{For simplicity the tunnelling amplitudes are assumed energy independent and identical in magnitude
between the different $m$-agglomerates $|t_m|=t$ to minimise the number of parameters. In real situations this may not be the case, but the formalism can be easily extended.}

A finite bias $V$ between the two edges will be considered and included in our formalism by gauge transforming the $m$-agglomerate tunnelling amplitudes 
with a time-dependent phase 
$t_{m}\to t_{m}e^{im\omega_0t}$. Here, $\omega_0=e^*V/\hbar$ is the Josephson resonance 
associated to the fundamental charge $e^*$.~\cite{Martin05} From the tunnelling Hamiltonian in
Eq.~(\ref{tun}) one can easily recover the current operator associated to the 
$m$-agglomerate contribution 
\be
I^{(m)}_B(t)=i me^* \left(t_{m}\  e^{im\omega_0t}\Psi_{\nu,+}^{(m)}(0,t){\Psi^{(m)}_{\nu,-}}^{\dagger}(0,t)-h.c.\right).
\ee
 
\subsection{Finite frequency noise}
A typical quantity to consider in order to investigate the noise properties of the QPC coupled to the resonant circuit is the non-symmetrized noise~\cite{Lesovik97, Gavish00, Aguado00} defined in terms of the backscattering current fluctuations
\be
S^{(m)}_{+}(\omega)=\frac{1}{2}\int^{+\infty}_{-\infty} dt \, e^{i \omega t}
 \langle \delta I^{(m)}_{B}(0) \delta I^{(m)}_{B}(t) \rangle. 
\ee
This quantity represents, for $\omega>0$, the noise emission of the system into the detector. The corresponding absorptive part is given by 
\be
S^{(m)}_{-}(\omega)=\frac{1}{2}\int^{+\infty}_{-\infty} dt \, e^{i \omega t}
\langle \delta I^{(m)}_{B}(t) \delta I^{(m)}_{B}(0) \rangle. 
\label{S_minus}
\ee
Here we have used the definition of back-scattering current fluctuation $\delta I_B^{(m)}= I^{(m)}_{B}-\langle I^{(m)}_{B}\rangle$ with the average $\langle...\rangle$ taken over the quantum statistical ensemble.  Note that the time reversal symmetry in the current-current correlator 
imposes that the emission noise at positive frequency must be identical to the absorption noise at negative 
frequencies,
\be
\label{TR}
S^{(m)}_{-}(\omega)=S^{(m)}_{+}(-\omega)\ .
\label{Plus_minus}
\ee 
The above quantities have been intensively investigated in recent years and different detection schemes were proposed.~\cite{Gavish00,Aguado00,Creux06,Zazunov07,Basset10,Basset12}

Historically, first approaches to the problem of the f.f. current fluctuations spectrum in mesoscopic systems were based on the so called symmetrized noise~\cite{Blanter00,Rogovin74, Chamon95, Chamon96,Carrega12} 
\beq
S^{(m)}_{sym}(\omega)&=& \frac{1}{2}\int^{+\infty}_{-\infty} d t \,e^{i \omega t}
\langle \{ \delta I^{(m)}_{B}(t), \delta I^{(m)}_{B}(0)\}\rangle
= S^{(m)}_{+}(\omega)+S^{(m)}_{-}(\omega)
\label{eq:s_sym}
\eeq
where $\{\cdot ,\cdot \}$ indicates the anticommutator. This appears the most 
natural definition of noise since the current operator, at different times, does not necessary commute.~\cite{Landaubook} 
This definition is also the correct one in the classical limit where the current at different time commute. However, this definition is reasonable
until $\hbar\omega\ll k_B T$, namely far from the quantum noise regime.~\cite{Rogovin74} 

In the limit $\hbar\omega\gg k_B T,e^*V$, one can expect to see the emergence of quantum effects. With nowadays technology,  at very short timescales (very high frequencies), comparable with natural energy scales $\hbar/k_BT$ or $\hbar/e^* V$, the current operator cannot be considered anymore a classical commuting quantity and the symmetrized (classical) definition could not be valid.~\cite{Lesovik97,Gavish00,Bednorz13,Aguado00} The detector of Fig.~\ref{fig_A} represents indeed a concrete measurement scheme to investigate current fluctuations at high frequencies. 

In the following we will concentrate on the regime where the QPC temperature is lower than the frequency (quantum limit) and the bias $\omega_0=(e^* V)/\hbar$ (shot noise limit). This will allow to clearly resolve the fractional qp contributions.

The measurable quantity, in the scheme of Fig.~\ref{fig_A}, is the output power proportional to the variation of the energy stored in the LC before and after the switching on of the LC-QPC coupling. We will indicate it as measured noise $S_{meas}$. At lowest perturbative order in the coupling $K\ll 1$ it can be expressed in terms of the non-symmetrized noise spectrum of the QPC.~\cite{Lesovik97, Gavish00, Chevallier10, Gavish04}
For an $m$-agglomerate contribution one has~\cite{Lesovik97, Bednorz13, Chevallier10, Gavish01}
\be
S^{(m)}_{meas}(\omega)= K \left\{S^{(m)}_{+}(\omega)+n_{B}(\omega) \left[ S^{(m)}_{+}(\omega)-S^{(m)}_{-}(\omega)\right]\right\}\ ,
\label{S_meas}
\ee
with (from now on $\hbar=1$ for notational convenience)
\be
n_{B}(\omega)= \frac{1}{e^{\beta_{c} \omega}-1}
\ee
the bosonic distribution function describing the equilibrium state of the LC detector
and $\beta_{c}=1/k_{B}T_{c}$ the detector inverse temperature.\footnote{The detector it is always assumed at equilibrium because the system is only weakly coupled $K\ll 1$ and the leakage towards the measurement chain is considered negligible.} Being perturbative, different contributions to the noise will be additive. Note that $T_c$ can be different from the system temperature $T$ since system and detector are coupled only electrically and their thermodynamic environment could be entirely different. As we will see, the tunability of the detector temperature is an important resource to address different regimes depending on the ratio  $T/T_c$. 

As a particular case we note that when the QPC is at equilibrium (no external bias $V=0$), and for $T=T_c$ any power can be exchanged between the system and the detector, i.e. it must be $S^{(m)}_{meas}(\omega)=0$.~\cite{Lesovik97,Bednorz13} This fact can be demonstrated by taking Eq.~(\ref{S_meas}), the time reversal symmetry in Eq.~(\ref{TR}), and considering the detailed-balance relation for the QPC at equilibrium, i.e.  $S^{(m)}_{+}(\omega)=e^{-\beta\omega}S^{(m)}_{+}(-\omega)$ with  $\beta=1/k_BT$. 

The expression for the measured noise in Eq. (\ref{S_meas}) formally satisfies the condition $S^{(m)}_{meas}(\omega)=S^{(m)}_{meas}(-\omega)$. This can be obtained by exploiting the condition (\ref{Plus_minus}) together with the identity $n_{B}(\omega)+n_{B}(-\omega)=-1$. Anyway in our analysis the LC detector resonant frequency is well-defined only for positive values. Therefore, in the paper all the formulas are valid only on the positive frequency branch.

Note that the expression in Eq.~(\ref{S_meas}) can be rewritten in a different way. Indeed the non-symmetrized noise correlators $S^{(m)}_{\pm}(\omega)$ are connected to the dissipative component of the differential ac conductance $G^{(m)}_{ac}(\omega)$~\cite{Safi08, Safi09, Safi10}, namely 
\be
S^{(m)}_{+}(\omega)-S^{(m)}_{-}(\omega)= - \omega\  \Re e{\left[G^{(m)}_{ac}(\omega)\right]}
\label{SGac}
\ee
where $\Re e{\left[...\right]}$ indicates the real part. The differential a.c. conductance corresponds to the 
response of the back-scattering current to an infinitesimal ac modulation $v(t)= v_{m} \cos{\omega t}$ 
in addition to the static bias voltage: $V(t)= V+v(t)$. This quantity can be calculated at finite bias $V$ 
in linear response~\cite{Safi08}
\be
\label{Gac}
G^{(m)}_{ac}(\omega)= \frac{1}{\omega}\int dt (e^{i\omega t}-1)\langle [ \delta I^{(m)}_{B}(t), \delta I^{(m)}_{B}(0)]\rangle
\ee
with the commutator  $[\cdot,\cdot]$ showing the explicit dependence  on the current-current correlator. Eq.~(\ref{SGac}) can be seen as a consequence of the non-equilibrium fluctuation dissipation theorem.~\cite{Rogovin74,Safi08, Safi09, Safi10}

Focusing on positive frequencies, we consider two important limits of Eq.~(\ref{S_meas})~\cite{Lesovik97}. For the detector quantum limit $k_{B} T_{c}\ll \omega $ one has
\be
\label{lowTc}
S^{(m)}_{meas}(\omega)\approx K S^{(m)}_{+}(\omega)+\mathcal{O}(e^{-\omega/k_BT_c}),
\ee  

leading to a direct measure of the emitted non-symmetrized noise correlator which, in this context, can be seen as a limiting case of the measured noise. In the opposite case $k_{B} T_{c}\gg \omega $ one has $n_B(\omega)\approx k_B T_c/\omega$ obtaining

\be
S^{(m)}_{meas}(\omega)\approx K \left\{S^{(m)}_{+}(\omega)-k_{B}T_{c} \Re e{\left[G^{(m)}_{ac}(\omega)\right]}\right\}
\label{GmeasApprox}
\ee
where we see a negative linear dependence of $S_{meas}$ on $T_c$. This is associated to the partial adsorption by the QPC of energy emitted by the detector through the dissipative component of the ac conductance. In other words tuning the temperature of the detector above (below) the frequency one can test mainly the absorptive (emissive) component of the current fluctuation. 
  
We recall that the total noise is given by the sum of the $m$-agglomerate contributions
\be
  S_{sym}(\omega)= \sum_{m} S^{(m)}_{sym}(\omega); \qquad S_{meas}(\omega)= \sum_{m} S^{(m)}_{meas}(\omega)
  \label{noise_sum}
  \ee
where the sum runs over all the relevant tunnelling excitations. Later we will also consider the derivatives of the noise with respect to the QPC bias, i.e. 
\be
\frac{\partial S_{sym}(\omega, \omega_{0})}{\partial \omega_{0}}; \qquad \frac{\partial S_{meas}(\omega, \omega_{0})}{\partial \omega_{0}}.
\ee
Before proceeding we would like to comment on the measurability of $S_{meas}(\omega,\omega_0)$.
At first sight one may wondering if it is possible to tune the system-detector coupling without introducing significant disturbance.
In experiments one usually keeps fixed $K$ investigating the excess noise power (for sake of clarity, from now on, we will indicate the dependence of the noise also on the bias $\omega_0$)
\be
S_{ex}(\omega,\omega_0)=S_{meas}(\omega,\omega_0)-S_{meas}(\omega,\omega_0=0)
\ee
which coincides with the difference of $S_{meas}$ between equilibrium and out-of-equilibrium conditions.\footnote{Note that it has been also demonstrated that $S_{ex}(\omega, \omega_0)$ is the best quantity to be amplified without introducing a relevant measurement distorsion.~\cite{Gavish00}}
However, in this work we mainly discuss $S_{meas}$, keeping fixed frequency $\omega$. One can easily obtain $S_{ex}$ shifting our results with $S_{meas}(\omega,0)$. Indeed, we think that a discussion of $S_{meas}$ is more transparent being directly connected to the non-symmetrized noise $S_+(\omega,\omega_0)$. Note that $S_{ex}(\omega,\omega_0)\equiv S_{meas}(\omega,\omega_0)$ if the system and detector have the same temperature ($T=T_c$) where indeed $S_{meas}(\omega,0)\equiv 0$.

We conclude this general section reminding the relation of the non-symmetrized emission noise with the rate of the quantum Hall QPC. At first order in the tunnelling amplitude, one can calculate the noise using the standard Keldysh formalism~\cite{Martin05}
\be
S^{(m)}_{+}(\omega, \omega_{0})= \frac{(m e^{*})^{2}}{2} \left[{\bf{\Gamma}}^{(m)}\left(-\omega+ m\omega_{0}\right)+{\bf{\Gamma}}^{(m)}\left(-\omega- m\omega_{0}\right)\right],
\label{S_plus_rate}
\ee
where we have introduced the QPC $m$-agglomerate tunnelling rate.~\cite{Gavish00} Note that from the previous equation one immediately observes the symmetry in the bias $S^{(m)}_{+}(\omega, \omega_{0})=S^{(m)}_{+}(\omega, -\omega_{0})$. 

Finally, by replacing Eq.~(\ref{S_plus_rate}) into Eq.~(\ref{eq:s_sym}) and Eq.~(\ref{SGac}) one has
\beq
\label{SsymRate}
S^{(m)}_{sym}(\omega)&=&\frac{(m e^{*})^{2}}{2}
\sum_{j,k=\pm}
{\bf{\Gamma}}^{(m)}\left(j\omega+k m\omega_{0}\right)\\
\label{GacRate}
 \Re e{\left[G^{(m)}_{ac}(\omega)\right]}&=&\frac{(m e^{*})^{2}}{2\omega}
 \sum_{j,k=\pm} j
{\bf{\Gamma}}^{(m)}\left(j\omega+k m\omega_{0}\right)
\eeq
 in full agreement with Ref.~\cite{Safi10}.

\subsection{Tunnelling rates}
The tunnelling rate ${\bf{\Gamma}}^{(m)}(E)$ introduced in the previous section is defined as
 
 \be
 \label{rate}
 {\bf{\Gamma}}^{(m)}(E) = |t_{m}|^{2} \int^{+\infty}_{-\infty} d \tau\  e^{i E t} \mathcal{G}^{<}_{m, -}(-t) \mathcal{G}^{>}_{m, +}(t),
 \ee
 with $\mathcal{G}^{>}_{m, \pm}(t)=\langle\Psi_{\nu,\pm}^{(m)}(t)\Psi_{\nu,\pm}^{(m)\dagger}(0) \rangle=(\mathcal{G}^{<}_{m, \pm}(-t))^*$ the greater/lesser correlation functions associated to the $m$-agglomerate operator for the edge $j= \pm$. For states belonging to the Laughlin sequence, with $\nu=1/(2n+1)$, this correlation function is~\cite{Ferraro08, Ferraro10}
 \be
 \label{Ggreater}
\mathcal{G}^{>}_{m,\pm}(t)=\frac{1}{2\pi a} \left[\mathcal{G}^>_{0,\pm}\left(\frac{t}{\beta},\beta\omega_c\right)\right]^{2\Delta_{\nu,c}^{(m)}}\!\!\!\!\!\!\!\!=\frac{1}{2\pi a} \left[\frac{|\Gamma(1+(\beta \omega_{c})^{-1}-i t /\beta )|^{2}}{\Gamma^{2}( 1+(\beta \omega_{c})^{-1}) (1\pm i \omega_{c} t)} \right]^{2\Delta_{\nu,c}^{(m)}}\!\!\!\!\!\!\!\!,  
\ee
where $\Gamma(x)$ is the  Euler gamma function and $\omega_{c}=v_{c}/a$ is the high frequency cut-off assumed as the greatest energy scale. The exponent depends on the scaling dimension of the charged mode operators $\Psi_{\nu,j}^{(m)}$. Note that $\mathcal{G}_{0,\pm}^>(t)$ represents the non-interacting 
two-point greater correlation function. For $\beta \omega_c,t/\beta\gg 1$ one gets
\be
\mathcal{G}^>_{0,\pm}\left(\frac{t}{\beta},\beta\omega_c\right)\approx\frac{\pi T}{\omega_c \sinh[\pi T(\mp t+i0^+)]}
\ee
that is the result of Eq.(\ref{Ggreater}) for $m=1$ (the electron) in the chiral fermi liquid where $\Delta_{1, c}^{(1)}=1/2$.

For the Laughlin sequence is $\Delta_{\nu,c}^{(m)} =m^2\nu/2$. The rate can be calculated at zero temperature for $E\ll\omega_c$
\be 
 {\bf{\Gamma}}^{(m)}(E) = \frac{|t_{m}|^{2}}{2\pi a^2}\left(\frac{E}{\omega_{c}} \right)^{4 \Delta_{\nu,c}^{(m)} } \frac{E^{-1}}{\Gamma(2\nu)} \theta(E)\propto \theta(E)E^{4 \Delta_{\nu,c}^{(m)}-1}
 \label{Rate_zeroT}
 \ee
with $\theta(x)$ the unit step function. This shows the typical power-law energy dependence of the interacting chiral Luttinger liquid.~\cite{Guinea95, Cuniberti96, Braggio00, Braggio01}

The tunnelling rates, in the case of composite edge states of the Jain sequence depend also on the neutral modes. The correlation function then becomes 
\be 
\mathcal{G}^{>}_{m, \pm}(t)= \frac{1}{2\pi a}\left[
\mathcal{G}_{0,\pm}\left(\frac{t}{\beta},\beta\omega_c\right)
\right]^{2\Delta_{\nu,c}^{(m)}}   
\left[\mathcal{G}_{0,\pm\textrm{\scriptsize sgn}(p)}\left(\frac{t}{\beta},\beta\omega_n\right)
\right]^{2\Delta_{\nu,n}^{(m)}}  
\label{single_due_quinti}
\ee
where $\omega_{n}=v_{n}/a$ is the neutral mode cut-off frequency. Usually $\omega_n\ll \omega_{c}$ (since $v_n\ll v_c$) and may be comparable with other energy scales (frequency, voltage, temperature).\footnote{The rate in the case of co-propagating or counter-propagating  the neutral mode is exactly the same if the QPC is point-like, i.e. described by Eq.~(\ref{tun}).} 
In the previous equation we have separated the contribution of charged ($\Delta_{\nu,c}^{(m)}$) and neutral ($\Delta_{\nu,n}^{(m)}$) modes. The exponents depend on the value of the filling factor and on the charge of the tunnelling excitations (see the discussion after Eq.(\ref{scaling_jain})).
In particular, for the most dominant $m$-agglomerate excitation, one finds the charge mode scaling $\Delta_{\nu,c}^{(m)}=(m^2/p^2)\nu/2$ and the neutral contribution, only present when $m$ is odd,  is $\Delta_{\nu,n}^{(m)}=1/4$ .

From these expressions, all the other quantities can be easily calculated using Eqs.~(\ref{S_plus_rate}-\ref{GacRate}). 

 It is worth to note that in the quantum limit $k_B T_c\ll \omega$ and in the shot noise limit, $\omega_{0}\gg k_{B}T$, for $m\omega_0\backsim \omega$ one has
 \be
S_{meas}^{(m)}(\omega,\omega_0)
\approx  S_{+}^{(m)}(\omega)\approx K\frac{(me^*)^2}{2}{\bf{\Gamma}}^{(m)}\left(-\omega+ m\omega_{0}\right).
 \ee

 Thus, $S_{meas}^{(m)}$ returns directly information on the tunnelling rate ${\bf{\Gamma}}^{(m)}(E)$.

 In particular, keeping fixed the LC frequency $\omega$ and changing the bias voltage $\omega_0$ 
 one can directly access its behaviour. We will see that, in the presence of more than one tunelling excitations, 
this is the most powerful tool in order to extract direct and unique information on rates. Indeed, different $m$-agglomerate contributions to $S_{meas}$ are directly proportional to the rates shifted in bias by $\omega_m=e^*V_m=\omega/m$. Therefore it is possible to investigate the rate behaviour 
of the different contributions separately, and to reconstruct the energy dependence of the tunnelling density of states associated to the different tunnelling processes. 

Note that the d.c. current cannot be helpful since all contributions are mixed. Also, other f.f. quantities such as the symmetrized noise $S_{sym}(\omega,\omega_0)$ (as we will better see later), 
could show signatures at bias $V_m$ associated to the resonances with $m$-agglomerates. However, even assuming to be able to build a setup able to measure it at high frequencies, it cannot convey such a clear signature due to the peculiar combination of rates it is composed, see Eq.~(\ref{SsymRate}). A similar argument limitation applies also  for the dissipative part of the differential conductance $G_{ac}(\omega, \omega_{0})$ (see Eq.~(\ref{GacRate})).

Therefore in the proposed setup, with frequency comparable with the bias $\omega_0$ and $k_BT,k_BT_c\ll\omega$ one can extract information on the tunnelling rates of different $m$-agglomerates using a single measurement at fixed frequency changing only the QPC bias. In the next section we will demonstrate that, for realistic value of physical parameters, this measurement is feasible. 
 
 These are the reasons why is crucial to consider f.f. detection scheme, 
  especially in perspective of the issue of identifying the coexistence of different excitations 
  with different charges in a quantum Hall fluid.  

 \section{Results and discussions}
 \label{results}
 In this section we present our main results on the f.f. noise detection for quantum Hall states in the Laughlin and Jain sequences.
We will focus on the dominant excitations in the tunnelling process analyzing their signatures in the f.f. symmetrized noise $S_{sym}$ as well as in the LC detector power output $S_{meas}$. We will discuss in detail the differences between these two quantities and which information they can convey.

Let us first discuss the specific values of the external parameters we have considered, chosen to be as close as possible to experimental values.  

The QPC temperature is kept the lowest possible (around $10$ mK) in order to investigate the so-called shot noise limit $k_BT\ll\omega_0$. The bias $\omega_0$ will be scanned around the LC frequency $\omega$ keeping it fixed. 
This procedure is, by far, the easier to be realised experimentally because, keeping fixed the frequency, one need to tune the impedance matching circuit to be in the appropriate regime only at the very beginning of the experiment. Impedance matching, at those frequency, is probably the most difficult challenge in such measurements. Furthermore, the majority of the previous theoretical literature concerning f.f. noise focussed on the frequency dependence of the spectral noise. In this work we consider the functional dependence of the f.f. noise on different quantities, observing features not easily visible with the conventional approach.

The frequency must be chosen to be the highest possible in order to minimize thermal effects, but still in a reasonable range (for nowadays technology) \emph{i.e.} between $5-10$ GHz (we choose $7.6$ GHz$\  \backsim 60$ mK). 
QPC and LC detector temperatures will be usually considered in the quantum limit 
($k_BT,k_BT_c\ll\omega$), where the contribution to the LC output power is strongly influenced by the emission part of the QPC noise spectrum.  In some cases larger values will be considered around the resonant frequency of the detector. 

We do not investigate higher detector temperatures since, in that case, mainly the adsorptive part of the noise spectrum will be probed. In that case the detection scheme can be more efficiently substituted by a direct measurement of the dissipative component of the differential conductance. We will also show extremely low temperatures for the QPC, {i.e.} $0.1$ mK, only with the aim of magnifying the QPC signatures.

It is worth mentioning that the symmetrized noise has been intensively investigated before. In the seminal paper Ref.~\cite{Chamon95} it has been recognized its strength in order to extract information on the Josephson resonances associated to fractionally charged excitations and their scaling. In particular it has been predicted a potentially singular behaviour in frequency around the Josephson resonance $m\omega_0$, for a generic $m$-agglomerate, that will be $|\omega-m\omega_0|^{4\Delta_\nu^{(m)}-1}$ with $\Delta_\nu^{(m)}$ the scaling dimension discussed in Eq.~(\ref{scaling_jain}).~\cite{Ferraro10,Ferraro12} We will see that a check of this behaviour in real experiments it may be difficult with actual technology. 
 
 \subsection{Laughlin sequence}
\begin{figure}[ht]
\centering
\includegraphics[scale=0.60]{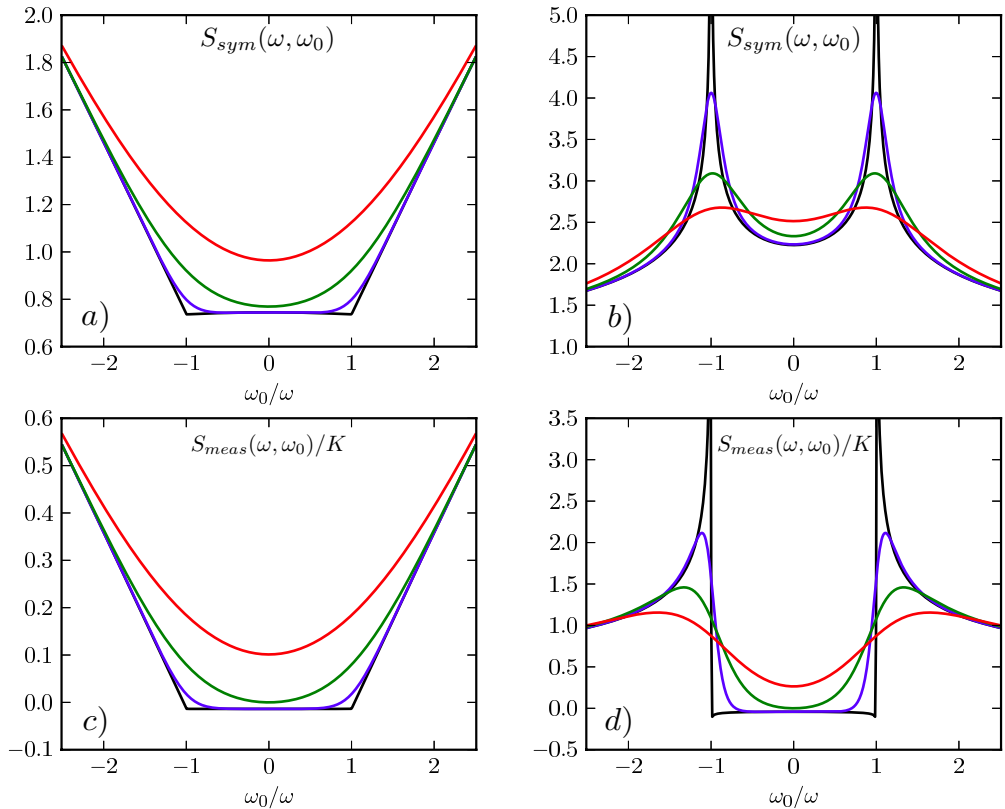}
\caption{$S_{sym}(\omega, \omega_{0})$ for a) $\nu=1$ and b) $\nu=1/3$ and $S_{meas}(\omega, \omega_{0})/K$ for c) $\nu=1$ and d) $\nu=1/3$ (in units of $S_0=e^{2} |t_{1}|^{2}/(2\pi \alpha)^2\omega_{c}$) as a function of $\omega_{0}/\omega$, with $\omega_0 = e^* V$. Temperatures are: $T=0.1$ mK (black), $T=5$ mK (blue), $T= 15$ mK (green) and $T=30$ mK (red). Other parameters are: $T_{c}=15$ mK, $\omega=7.9$ GHz ($60$ mK), $\omega_{c}=660$ GHz ($5$ K) and $g_c=1$.}
\label{fig_B}
\end{figure}

In the Laughlin case, $\nu=1/(2n + 1)$ the dominant tunnelling contribution is given by the single-qp with charge $e^{*}=\nu e$. 

For these excitations we will analyze the f.f. symmetrized noise $S_{sym}(\omega,\omega_0)$, comparing it with the measured noise $S_{meas}(\omega,\omega_0)$. Figures \ref{fig_B} a) and b) show $S_{sym}$ as a function of the Josephson frequency $\omega_0 = e^* V$, for different temperatures at $\nu=1$ and $\nu=1/3$ respectively. 
  
Note that, at integer filling factor $\nu=1$ (non-interacting case), is the electron which tunnels through the QPC. The curve, at extremely low temperature (black line), is flat in the interval $|\omega_{0}/\omega|\leq 1$ and increases linearly outside this range, in full agreement with the expected behaviour for a normal Fermi liquid~\cite{Blanter00, Rogovin74, Dahm69, Jacob83, Schoen85, Zakka07}. By increasing temperatures one observes a progressive rounding and vertical shift of the curves.~\cite{Blanter00,Zakka07}
 
 Different is the situation for $\nu=1/3$ where two evident peaks appear at $|\omega_{0}/\omega|= 1$. They are clear fingerprints of the interacting $\nu=1/3$ state associated to the presence of the single-qp excitation $e^*$.~\cite{Chamon95,Safi08} The bias dependence of the symmetrized noise near the resonances is given by $|\omega_0-\omega|^{4\Delta_{1/3}^{(1)}-1}$, with $\Delta_{1/3}^{(1)}=1/6$ the single-qp scaling dimension. The possible resonances are strongly washed out by thermal effects.
 
Note that the low temperature curve for $S_{sym}$ start from a non-zero positive value. This  is related to the fact that $S_{sym}$, when $k_BT\ll\omega$, shows the contributions from the quantum noise associated to the ground state of the system.  It has been longly discussed if those contributions are effectively measurable since one cannot extract any energy from the ground state of a system with a passive detector.~\cite{Gavish00,Bednorz13} 

The $S_{meas}$, defined in Eq. (\ref{S_meas}), is a concrete example of a different quantity related to the current noise with a well defined measurement prescription. In Figs. \ref{fig_B} c) and d) we represent $S_{meas}$ for $\nu=1$ and $\nu=1/3$ respectively, varying the temperature of the system and keeping fixed the detector temperature $T_{c}$. 

For $\nu=1$, $S_{meas}$ is qualitatively similar to $S_{sym}$ (see Fig. \ref{fig_B} a)  ).
The curves are almost identical apart a constant shift. This is confirmed by a simple analytical calculation. Recalling the zero temperature behaviour of the electron tunnelling rate 
(see Eq.(\ref{Rate_zeroT})) for a chiral Fermi liquid ($\nu=1$),
${\bf\Gamma}^{(1)}(E)\propto \theta(E) E$, one finds from Eq. (\ref{SsymRate}) for $\omega_{0}>0$,
\be
S_{sym}(\omega,\omega_0)= 2\tilde{S}_0\left[\theta(\omega_0-\omega)\omega_0+\theta(\omega-\omega_0)\omega\right]/\omega_c
\ee

with $\tilde{S}_0=\frac{e^{2}}{2}\frac{|t_{1}|^{2}}{2\pi \alpha^2} \frac{1}{\omega_{c}}$.  On the other hand, assuming the detector in the quantum limit ($k_BT_c\ll\omega$), from Eq. (\ref{lowTc}) and Eq. (\ref{S_plus_rate}), one gets ($\omega_{0}>0$)
\be
S_{meas}(\omega,\omega_0)\approx K S_{+}(\omega,\omega_0)=K \tilde{S}_0 \theta(\omega_0-\omega)(\omega_0-\omega)/\omega_c\ .
\ee
Thus we have
\be
\label{relation}
S_{meas}(\omega,\omega_0)\approx \frac{K}{2}\left(S_{sym}(\omega,\omega_0)-2\tilde{S}_0\frac{\omega}{\omega_c}\right).
\ee
This result is tightly connected to the linear dependence on energy of the tunnelling rate typical of a non-interacting electron systems. 

Note that, in the considered limits, the $S_{meas}$ curve starts from almost zero instead of a finite value reflecting the fact that, in 
this resonant measurement scheme, the ground state fluctuations cannot appear by definition.~\cite{Lesovik97}
 As before, thermal effects lead to a rounding and a shifting up of the curves.
 
 \begin{figure}[ht]
\centering
\includegraphics[scale=0.60]{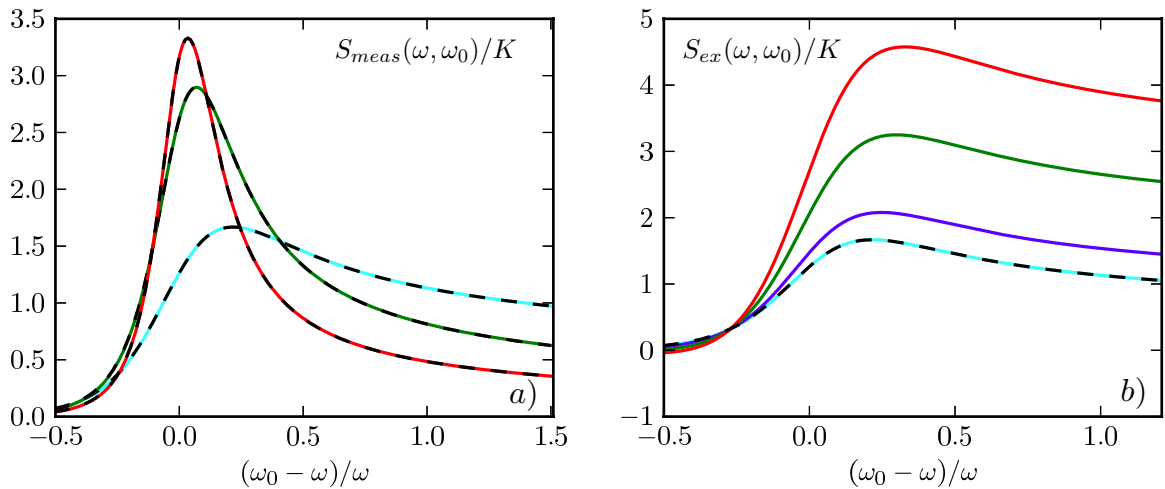}
\caption{Panel a) comparison between $S_{meas}(\omega, \omega_{0})/K$ (in $S_0$ units) (coloured lines) at finite temperature $T=T_{c}=10$ mK and the finite temperature rate ${\bf\Gamma}^{(1)}(E)$ (black dashed line), rescaled by a constant factor, at different filling factors: $\nu=1/3$ (cyan), $\nu=1/5$ (green) and $\nu=1/7$ (red).  Panel b) $S_{ex}(\omega,\omega_0)/K$ (in $S_0$ units) for $\nu=1/3$ at $T=10$mK and for different values of $T_c= 10$mK (cyan), $30$mK (blue), $60$mK (green), $90$mK (red).
Other parameters are $\omega=7.9$ GHz ($60$ mK), $\omega_{c}=660$ GHz ($5$ K).}
\label{fig_Rate}
\end{figure}

Different is the interacting case ($\nu=1/3$) shown in Fig. \ref{fig_B} d). 
The curve for $S_{meas}$ at the lowest temperature starts close to zero at $\omega_{0}/\omega=0$ and jumps abruptly at $|\omega_{0}/\omega|\approx 1$ reaching two peaks with a structure similar to $S_{sym}$ in Fig \ref{fig_B} b) but with a shape totally different (asymmetric) and more resolved with respect to bias. 
 This difference is due to the fact that the symmetrized noise around the LC frequency comes from a combination of forward and backward rate, while $S_{meas}$ for $k_{B}T_{c}\ll \omega$ and $k_{B}T\ll \omega_{0}$ \emph{measures} the tunnelling rate of the tunnelling excitations (see below). By increasing the temperature this peaked structure progressively disappears as expected. The maxima become less pronounced and move to higher values of $|\omega_{0}/\omega|$ due to thermal corrections, however the jumps in the curves remains a clear signature of the single-qp contribution to the noise. 
 
The fact that $S_{meas}$ returns directly information on the rate can be verified by looking at Fig.~\ref{fig_Rate} a). Here it is shown the behaviour of $S_{meas}/K$ (with $T=T_c$) as a function of the bias $(\omega_0-\omega)/\omega$ around the LC frequency for different values of the filling factors, i.e. different scaling dimension of the single-qp operators.

One can see that $S_{meas}$ reproduces the rate (black dashed lines) (see Eq. (\ref{rate})) with the same scaling dimension. Its fidelity is so good that one can extract the scaling dimension by directly fitting this quantity at finite temperature. It is worth to note that we choose the condition $T=T_c$ because in such case $S_{meas}\equiv S_{ex}>0$, a necessary condition since the transition probability is positive. 

One can carry out this measurement also for $T\neq T_c$ considering $S_{ex}$ (see Fig. \ref{fig_Rate} b)). Looking at this figure is immediately clear that, increasing $T_c$, the connection to the rate fails since the contribution of the absorption part strongly modifies the profile of $S_{ex}$ with respect to the predicted lineshape of the rate (black dashed lines), which, of course, does not depend on $T_{c}$.

 In conclusion, at temperatures $k_BT_c, k_BT\ll\omega$, we can directly compare $S_{meas}$ with the rate extracting information about the scaling dimension of the tunnelling excitation even without the necessity to go at extremely low temperatures.

\begin{figure}[ht]
\centering
\includegraphics[scale=0.60]{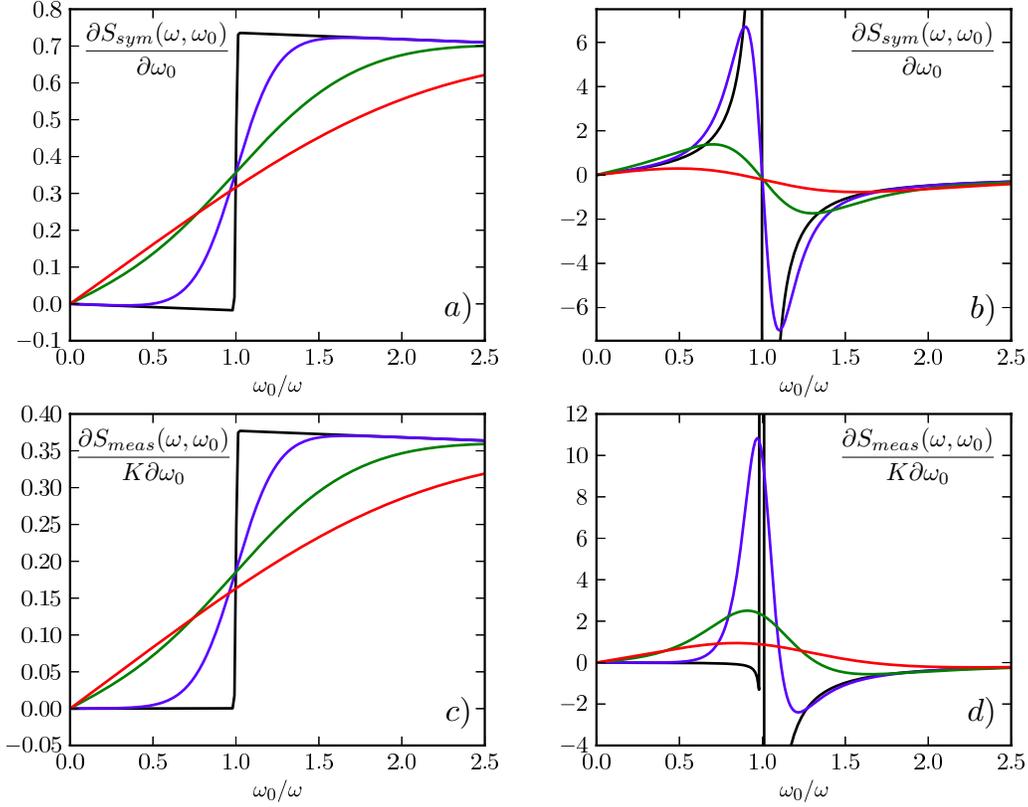}
\caption{$\partial_{\omega_0} S_{sym}(\omega, \omega_{0})$ for a) $\nu=1$ and b) $\nu=1/3$ and $\partial_{\omega_0} S_{meas}(\omega, \omega_{0})/K$ for c) $\nu=1$ and d) $\nu=1/3$ (in units of $S_0/\omega$) and as a function of $\omega_{0}/\omega$, with $\omega_0=e^*V$. Temperatures are: $T=0.1$ mK (black), $T=5$ mK (blue), $T= 15$ mK (green) and $T=30$ mK (red). Other parameters are: $T_{c}=15$ mK, $\omega=7.9$ GHz ($60$ mK), $\omega_{c}=660$ GHz ($5$ K).}
\label{fig_C}
\end{figure}

The different behaviours between non-interacting ($\nu=1$) and the interacting ($\nu=1/3$) case 
can be better appreciated by considering $\partial_{\omega_0}S_{sym}$ and $\partial_{\omega_0}S_{meas}$ as shown in Fig. \ref{fig_C} as a function of the bias $\omega_{0}>0$. Note that for $\omega_0<0$ one can easily use the property $\partial_{\omega_0}S_{meas}(\omega,\omega_0)=-\partial_{\omega_0}S_{meas}(\omega,-\omega_0)$. 

Fig. \ref{fig_C} a) shows $\partial_{\omega_0}S_{sym}$ at $\nu=1$ (non-interacting case) which at very low temperature (black) presents a step structure related to the Fermi liquid nature of this state. \footnote{The steps are slightly tilted due to the finite cut-off energy $\omega_{c}$ considered here.} By increasing temperature this structure progressively linearizes as one would expect when $\omega_0\backsim k_B T$ (red). 

For $\nu=1/3$ in Fig. \ref{fig_C} b), the derivative has zeros at $|\omega_0/\omega|=1$, signature of the non Fermi liquid nature of the excitations and revealing the position of the Josephson resonance connected to the single-qp excitation.

Comparing $\partial_{\omega_0}S_{meas}$ in the non-interacting case (Fig. \ref{fig_C} c)) with Fig. \ref{fig_C} a), one immediately sees the relation $\partial_{\omega_0} S_{meas}\approx(K/2) \partial_{\omega_0} S_{sym}$ as an obvious consequence of Eq.(\ref{relation}) generally valid at low temperatures.  

For the interacting case $\nu=1/3$, $\partial_{\omega_0}S_{meas}$ in Fig. \ref{fig_C} d) shows a pronounced peak around $\omega_{0}/\omega=1$ related, as stated before, to the single-qp contribution and appears strongly asymmetric with respect to $|\omega_{0}/\omega|=1$. The comparison with Fig. \ref{fig_C} b) confirms that, in the interacting case, there is not a simple relation connecting $S_{sym}$ and $S_{meas}$. 

\begin{figure}[ht]
\centering
\includegraphics[scale=0.60]{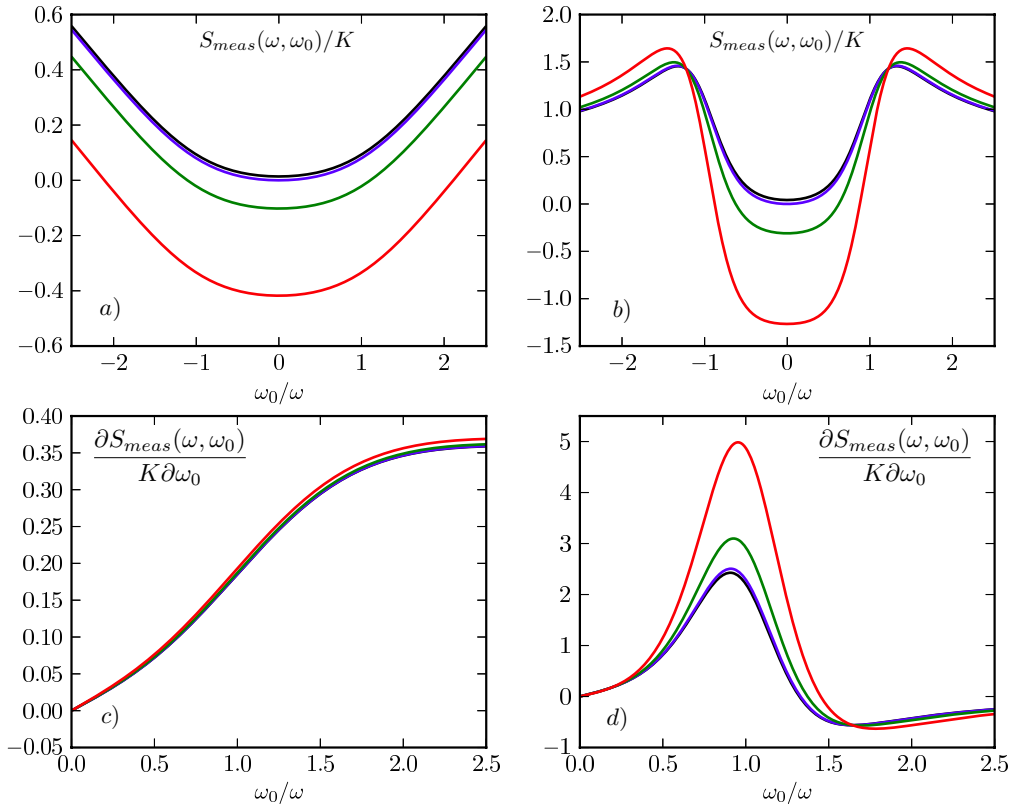}
\caption{Top panels: $S_{meas}(\omega, \omega_{0})/K$ (in units of $S_0$) as a function of $\omega_{0}/\omega$ for a) $\nu=1$ and b) $\nu=1/3$ at different values of the LC circuit temperature $T_{c}=5$ mK (black), $T_{c}=15$ mK (blue), $T_{c}= 30$ mK (green) and $T_{c}=60$ mK (red). Bottom panels:
$\partial_{\omega_0} S_{meas}(\omega, \omega_{0})/K$ (in units of $S_0/\omega$) as a function of $\omega_{0}/\omega>0$ for c) $\nu=1$ and d) $\nu=1/3$ at different $T_c$ (same colour code of top panels). Other parameters are: $T=15$ mK, $\omega=7.9$ GHz ($60$ mK) and $\omega_{c}=660$ GHz ($5$ K).}
\label{fig_DE}
\end{figure}

We now comment  on the effect of the detector temperature on $S_{meas}$. To this end we fix the temperature of the QPC system at $T=15$~mK.
 The behaviour of the measured noise and its bias derivative are shown in Figs. \ref{fig_DE} for different values of the detector temperature $T_c$ . 
 
 For $\nu=1$ the curves for $S_{meas}$ in Fig. \ref{fig_DE} a) start very close to zero at $T_{c}=5$ mK (black curve)  and are progressively shifted to negative values by increasing $T_{c}$ with the shift proportional to $T-T_{c}$ (see Ref.~\cite{Lesovik97}). This is due to the increasing dominance of the absorptive part, proportional to the dissipative component of $G_{ac}$ (see Eq. (\ref{GmeasApprox})), combined with the linearity of the system. Indeed, for a Fermi liquid, $G_{ac}$ is energy independent, leading to the observed rigid shift of all curves. This is confirmed by the bias derivative in Fig. \ref{fig_DE} c) where, as expected, all the curves coincide. 

More involved is the situation for $\nu=1/3$. In this case $S_{meas}$ in Fig. \ref{fig_DE} b), at $\omega_{0}/\omega \approx 0$ and at the lowest detector temperature (black line), is very close to zero  and then jumps abruptly for $|\omega_{0}/\omega|\approx 1$. By increasing $T_{c}$ the central region becomes negative, while the tails $|\omega_0/\omega|\geq 1$ are correspondingly enhanced. This behaviour is due to the presence of both absorption and emission contribution in the measured noise.  Indeed, for $|\omega_0/\omega|\ll1$ and sufficiently low QPC temperature ($|\omega-\omega_0|\gg k_B T$), the system cannot excite the detector mode $\omega$ absorbing only energy. In this regime, $S_{meas}$ decreases. Conversely, when $|\omega_0/\omega|\approx 1$, the system can emit photons into the detector with the best efficiency near the resonance (where the rate for $\nu=1/3$ is peaked at small $T$).  The combined effect of these two phenomena produces an enhancement of the jump around the resonance $|\omega_{0}/\omega| \approx 1$ by increasing $T_{c}$, as shown in Fig. \ref{fig_DE} b). Analogously there is a more pronounced peak in $\partial_{\omega_{0}} S_{meas}$ (Fig. \ref{fig_DE} d)) just at $\omega_{0}/\omega=1$ when fractional qp resonance is detected. 

The above behaviour suggests that increasing $T_c$ one could better identify the presence of these resonances. In any case caution is necessary in the regime of higher $T_{c}$ due to an increasing of the outgoing signal from the detector backgroung that can hide the QPC contribution and the growing importance of higher order effects in the coupling between system and detector that are neglected in our discussion.

Note that in the limit $k_{{\rm B}}T_c \gg \hbar \omega$ the measured noise is essentially governed by 
$\Re e[G_{ac}]$ as shown in Eq. (\ref{GmeasApprox}). In this case the absorption properties of the system are mainly detected.

\subsection{Jain sequence}
\begin{figure}[ht]
\centering
\includegraphics[scale=0.60]{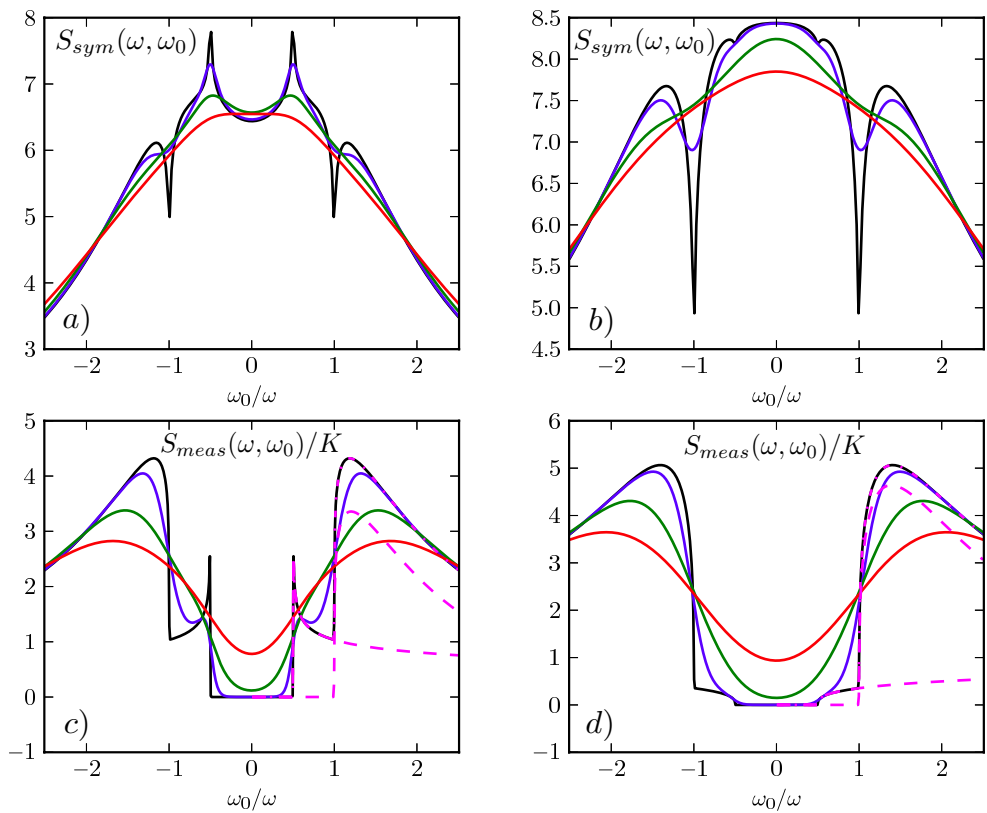}
\caption{$S_{sym}(\omega,\omega_0)$ for a) $\nu=2/5$ and b) $\nu=2/3$ and $S_{meas}(\omega,\omega_0)/K$ for c) $\nu=2/5$ and d) $\nu=2/3$. All quantities are in units of $S_0$ and as a function of $\omega_{0}/\omega$. Temperatures are: $T=0.1$ mK (black), $T=5$ mK (blue), $T= 15$ mK (green) and $T=30$ mK (red). Other parameters are: $\omega = 7.9$~GHz ($60$~mK), $T_{c}=0.1$ mK, $\omega_{n}=6.6$ GHz ($50$ mK), $\omega_{c}=660$ GHz ($5$ K) and $|t_{2}|^{2}/|t_{1}|^2=1$. The dashed lines correspond to the rate contributions of the $2$-agglomerate and the single-qp for $T=0.1$ mK. They are calculated separately and fitted only by changing their prefactor. The dashed-dotted line is the sum of the two contribution and returns exactly the behaviour of $S_{meas}$.}
\label{fig_F}
\end{figure}

We now discuss the case of composite filling factors focusing on $\nu=2/5$ and $\nu=2/3$.

As stated before, for $\nu=2/5$, the two dominant excitations are the single-qp 
$\Psi_{2/5}^{(1)}$ and the $2$-agglomerate $\Psi_{2/5}^{(2)}$. At zero frequency a crossover in the relevance of these two excitations has been observed and theoretical calculations confirm this scenario.~\cite{Wen95,Chung03,Ferraro08, Ferraro10} In the following, without loss of generality, we will assume $|t_{2}|^{2}/|t_{1}|^2=1$ as the ratio between the single-qp and $2$-agglomerate tunnelling amplitudes. 

Fig. \ref{fig_F} a) shows $S_{sym}$ as a function of the bias voltage $\omega_{0}/\omega$. The neutral mode cut-off $\omega_{n}\ll\omega_c$ has been chosen in a range compatible with previous fitting of real experimental data.~\cite{Ferraro08,Ferraro10,Carrega12}.

At extremely low temperature $T=0.1$ mK it is still possible to observe two little peaks at $|\omega_{0}/\omega|=1/2$ related to the presence of the $2$-agglomerate and two dips at $|\omega_{0}/\omega|=1$ due to the single-qp.
These different behaviours are due to the different bias dependence of $S_{sym}$ around $\omega_m=e^*V_m=\omega/m$: $S_{sym}\propto |\omega_0-\omega_m|^{4\Delta_\nu^{(m)}-1}$ showing peaks (dips) depending on the scaling dimension $\Delta_\nu^{(m)}<1/2$ ($\Delta_\nu^{(m)}>1/2$).~\cite{Chamon95,Chamon96} Increasing the QPC temperature these features are rapidly washed out. 

A similar analysis can be done for $\nu=2/3$ where the single-qp and the $2$-agglomerate are equally dominant in the tunnelling process.~\cite{Ferraro10} The behaviour of $S_{sym}$ is presented in Fig. \ref{fig_F} b), as a function of $\omega_{0}/\omega$, the scaling dimensions of the dominant single-qp and $2$-agglomerate operators are different from the $\nu=2/5$ case justifying the different behaviour. Here, the dip structure at $|\omega_0/\omega|=1$, associated to the single-qp is evident at very low temperature, but only a slight dip due to the $2$-agglomerate can be identified at $|\omega_{0}/\omega|=1/2$.  At higher temperatures such little features are washed out and apparently only the single-qp contribution can be detected. 
 
For what it concerns the measured noise $S_{meas}$ in Fig. \ref{fig_F} c) ($\nu=2/5$) one observes, at $T=0.1$ mK, a very flat curve for $|\omega_{0}/\omega|<1/2$ corresponding to the regime where the QPC is only absorptive. Abrupt jumps associated to the $2$-agglomerate and to the single-qp are present for $|\omega_{0}/\omega|\approx 1/2$ and $|\omega_{0}/\omega|\approx 1$ respectively.

We would like now to comment on the connection between $S_{meas}$ and the rate at $T=T_{c}$. One can easily see how the lineshape of $S_{meas}$ is directly connected to the contribution of the tunnelling rates of the single-qp ${\bf\Gamma}^{(1)}(E)$ and the $2$-agglomerate ${\bf\Gamma}^{(2)}(E)$
\be
\label{shift}
S_{meas}(\omega,\omega_0)\approx \alpha_1 {\bf\Gamma}^{(1)}(\omega_0-\omega)+\alpha_2 {\bf\Gamma}^{(2)}(2\omega_0-\omega)
\ee
with $\alpha_1$ and $\alpha_2$ constant prefactors. A fitting is shown in the Fig. \ref{fig_F} c) for the black curve ($T=T_c$), having indicated with dashed lines separately the contribution of the two rates and with dashed-dotted lines the sum of the two. Note that we considered an extremely low temperature $T=0.1$ mK just to make the signature of the rates more visible, however the same analysis may be repeated for higher temperatures satisfying the constraint $k_BT,k_BT_c\ll \omega$. 

To validate the previous discussion we consider $\nu=2/3$ where different scaling dimension should be observable for the rate of the $2$-agglomerate. In Fig. \ref{fig_F} d) we see indeed around the bias resonance $|\omega_0/\omega|\approx1/2$ a behaviour of $S_{meas}$ without peaks. This signals a different energy behaviour of the correspondent rate in comparison to the agglomerate for the $\nu=2/5$ case. The dashed and dashed-dotted lines show the fit of Eq.(\ref{shift}) for the present case demonstrating again how the proposed measurement is able to extract the energy dependence associated to the different tunnelling excitations.

Thermal effects rapidly hide the peaked structure smoothening the curves but, hopefully, keeping still enough information to reconstruct the scaling dimension of the tunnelling operators. Note that, for $|t_{2}|^{2}/|t_{1}|^2 \lessgtr 1$ single-qp and $2$-agglomerate may be less or more visible with respect to each other, however, due to the shift in bias, it is still possible to identify even small signatures. 

\begin{figure}[ht]
\centering
\includegraphics[scale=0.60]{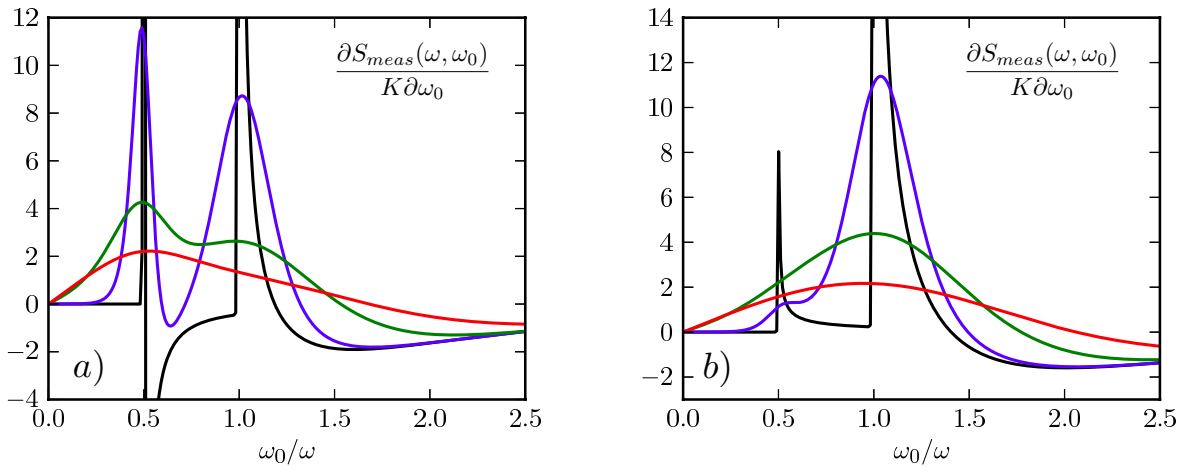}
\caption{$\partial_{\omega_0} S_{meas}(\omega, \omega_{0})/K$ (in units of $S_0/\omega$) at a) $\nu=2/5$ and b) $\nu=2/3$ as a function of $\omega_{0}/\omega>0$, with $\omega_0 = e^* V$. Temperatures are: $T=0.1$ mK (black), $T=5$ mK (blue), $T= 15$ mK (green), $T=30$ mK (red). Other parameters are: $\omega= 7.9$~GHz ($60$~mK), $T_{c}=15$ mK, $\omega_{n}=6.6$ GHz ($50$ mK), $\omega_{c}=660$ GHz ($5$ K) and $|t_{2}|^{2}/|t_{1}|^2=1$.}
\label{fig_G}
\end{figure}

The possibility to shift in the bias the signatures (peaks or dips) induced by the energy dependence of the rates is one of the most interesting results of this measurement protocol. This demonstrates that the resonant detector gives unique resource to separate the contribution of different charges allowing to reconstruct the fundamental properties of the tunnelling excitations: the fractional charges and their scaling dimension. In addition, this process may be done at finite (measurable) temperature for reasonable values of the LC detector circuit frequency $\omega$.

\begin{figure}[ht]
\centering
\includegraphics[scale=0.60]{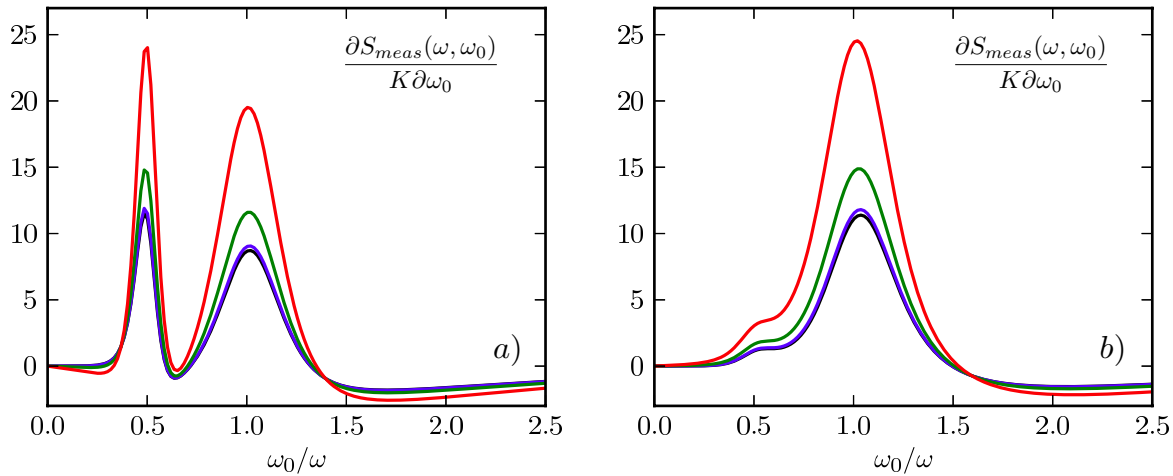}
\caption{$K^{-1}\partial S_{meas}(\omega, \omega_{0})/\partial \omega_{0}$ (in units of $S_0/\omega$) as a function of $\omega_{0}/\omega>0$ for a) $\nu=2/5$ and b) $\nu=2/3$ b) at different values of the LC circuit temperature: $T_{c}=5$ mK (black), $T_{c}=15$ mK (blue), $T_{c}= 30$ mK (green), $T_{c}=60$ mK (red). Other parameters are: $T=5$ mK, $\omega_{n}=6.6$ GHz ($50$ mK), $\omega=7.9$ GHz ($60$ mK), $\omega_{c}=660$ GHz ($5$ K) and $|t_{2}|^{2}/|t_{1}|^2=1$.}
\label{fig_J}
\end{figure}

This picture is confirmed by looking at the measured noise bias derivative in Fig. \ref{fig_G} a) and b). Here, peaks at $\omega_{0}/\omega=1/2, 1$ are visible for low enough temperatures. It is worth to note that, despite the difference in energy dependence of the tunnelling rate in the $2$-agglomerate for $\nu=2/5$ and $\nu=2/3$, the presence  the $2$-agglomerate in the bias scan is always signaled by a peak at sufficiently low temperatures.

Increasing $T_{c}$ at fixed system temperature $T$ (Fig. \ref{fig_J}) leads to a loss of resolution of the rate spectroscopy but it increases the sensibility in the detection of different $m$-agglomerates contributions. Obviously the enhancing of the sensibility with increasing $T_c$ is limited by the same general constraint on energy already discussed for the Laughlin case. 

As a final remark we would like to point out that the considered setup, with a similar detector scheme, can be useful also in order to investigate the photo-assisted noise spectra in the fractional Hall regime.~\cite{Chevallier10}

\section{Conclusions}
We have investigated finite frequency noise properties of FQHE states both in the Laughlin and the Jain sequences. Using a resonant detector coupled with a QPC in the weak back-scattering regime, we have studied the charge and the scaling dimension of the fractional tunnelling excitations. We have discussed in detail the difference between the ``measured noise'' $S_{meas}(\omega,\omega_0)$ and the f.f. symmetrized noise $S_{sym}(\omega,\omega_0)$ varying the bias $\omega_0=e^*V$ at fixed LC detector frequency $\omega$.
Taking the temperature of the detector in the quantum limit $\hbar\omega\gg k_B T_c$ and the QPC in the shot noise regime $e^*V\gg k_BT$, we demonstrated the possibility to detect the presence of different multiple qps excitations in the composite edge case. This regime is driven by the emission component of the measured noise allowing to  directly probe the tunnelling density of state associated to different excitations. 
We have shown that, increasing the detector temperature, one can improve the sensibility of the measurement and detects fractional qp excitations by testing the adsorptive component of the QPC spectrum.

\section*{Acknowledgements}
We thank C. Wahl, J. Rech, T. Jonckheere, T. Martin, F. Portier, and N. Magnoli for useful discussions. We acknowledge the support of the EU-FP7 via ITN-2008-234970 NANOCTM and MIUR-FIRB2012 - Project HybridNanoDev (Grant  No.RBFR1236VV).

\end{document}